# Plant cells-based biological matrix composites

Eleftheria Roumeli[1], Luca Bonanomi[1†], Rodinde Hendrickx[1†], Katherine Rinaldi[2], Chiara Daraio[1*]

[1]Division of Engineering and Applied Science, California Institute of Technology, Pasadena, CA 91125, USA.

[2]Division of Chemistry and Chemical Engineering, California Institute of Technology, Pasadena, CA 91125, USA.

*Correspondence to: daraio@caltech.edu.

† These authors contributed equally to this work.

The global increase in materials consumption calls for innovative materials, with tailored performance and multi-functionality, that are environmentally sustainable. Composites from renewable resources offer solutions to fulfil these demands but have so far been dominated by hybrid petrochemicals-based matrixes reinforced by natural fillers. Here, we present biological matrix composites with properties comparable to wood and commercial polymers. The biocomposites are obtained from cultured, undifferentiated plant cells, dehydrated and compressed under controlled conditions, forming a lamellar microstructure. Their stiffness and strength surpass that of commercial plastics of similar density, like polystyrene, and low-density polyethylene, while being entirely biodegradable. The properties can be further tuned varying the fabrication process. For example, filler particles can be integrated during fabrication, to vary the mechanical response or introduce new functionalities.

**One Sentence Summary:** We create natural, biodegradable composites from plant cells with properties akin to commercial plastics of similar density.

Polymer composites are amongst the most widely produced materials (*1*). However, their production and after-life use pose considerable environmental challenges (*2*). Most of the produced waste is disposed of in landfills or incinerated (*3*). Composites components made of sustainable or renewable resources, aka. biocomposites, offer promising solutions towards more sustainable products (*2*). Although the majority of biocomposites still contain petroleum-derived plastics as the main matrix material (*4*, *5*), research increasingly focuses on bio-derived or renewable matrix materials (*2*, *6*, *7*). Current challenges for fully bio-derived composites include balancing production costs with performance, improving durability and assessing the environmental impact of manufacturing processing and post-use strategies (*2*).

Engineered living materials (ELM) use living matter to fabricate and assemble the matrix components (*8*, *9*). Examples include materials derived from yeast fermentation in the presence of carbon nanotubes or graphene, which combine electrical and optical properties derived from the synthetic fillers, and self-healing properties from the living cells (*10*, *11*). Materials that combine living fungi or plant cells with carbon nanotubes are structurally stable and combine electrical conductivity (*12*) and temperature sensitivity (*13*). Mycelium materials already reached the market for protective packaging, insulation, and acoustic panels (*14–16*). The combination of wood particles, mycelium, and cellulose nanofibrils (CNF) resulted in composites with mechanical properties superior to all-mycelium materials (*17*). However, the main drawback of all existing biocomposites is their relatively low mechanical performance that renders them unsuitable for engineering and structural applications.

Plant materials demonstrate an impressive range of mechanical properties. Their stiffness and bending strength, for example, can vary over three orders of magnitude (*18*). This remarkable range depends on the native composition of the plant cell walls, on the arrangement of the different components within the cell wall (see Supplementary Information), and on the hierarchical organization of the cells at the microscale. Recently, chemical and/or thermo-mechanical post-processing of natural wood has been adopted to create high-performance materials, with properties comparable to steel, ceramics and insulating foams (*19–21*). However, wood processing relies on deforestation, transportation to treatment plants and harsh chemical treatments, which are not environmentally friendly.

Here we describe a new class of biocomposites based on cultured and dehydrated plant cells. Our materials retain the native plant cell wall composition naturally secreted by growing plant cells, to achieve mechanical performance comparable to structural and engineered woods, and polymers. We use undifferentiated, tobacco cells as a model system. We characterize the microstructure, composition and mechanical properties of the produced materials and show that the incorporation of filler additives allows to improve the material's performance and expand their functionalities, for example creating magnetic and electrically conductive materials.

We harvest plant cells from a suspension culture and compress them in a permeable mold, to achieve a densified dehydrated structure (See Fig. 1A and Materials & Methods). During compression, water diffuses through the plant cell wall and the cell volume is gradually reduced. When the cells reach a dry state, corresponding to approximately a 98% weight loss, the resulting material consists only of a lamellar stack of compacted cell walls. Cross-section scanning electron microscopy (SEM) of the resulting material (Fig. 1B) illustrates the obtained microstructure. We compare it to natural wood (walnut, Fig. 1C), commercial medium density fiberboard (MDF, Fig. 1D), and plywood (Fig. 1E). Our material is structurally similar to plywood and MFD, which are compressed wood composites bound together with polymer adhesives.

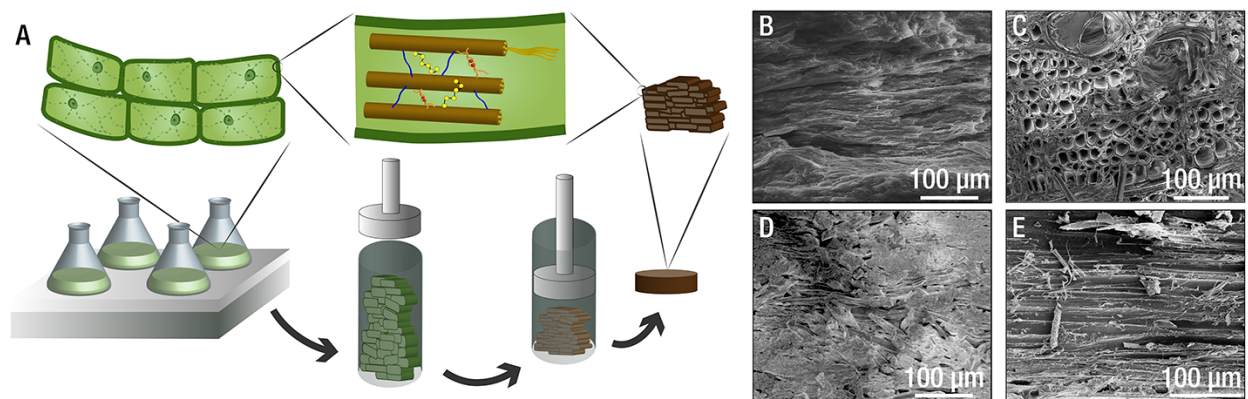

**Fig. 1. (A)** Schematic of the fabrication method. Plant cells are cultured, harvested and subjected to a controlled compression and dehydration, resulting in a lamellar architecture when dry. SEM cross-sectional views of the microstructure of **(B)** the biocomposite, **(C)** walnut, **(D)** MDF and **(E)** plywood at the same magnification.

We characterize the cell wall in living plant cells extracted from suspension cultures (Fig. 2A). Optical and laser microscopy shows that viable cells are elongated, with a mean length of 170±60 μm, a mean width of 45±10 μm, and are surrounded by a thin primary cell wall containing cellulose, pectin and phenolic compounds (Fig. 2B-D) (see Materials & Methods). Raman spectroscopy of living cells (Fig. 2E) reveals the predominant vibrations of cellulose, hemicelluloses, pectin, and the lignin precursors coniferyl alcohol and coniferaldehyde (*22*, *23*).

Compositional analysis of the dry material shows that it is composed of 15% cellulose, 20% hemicelluloses, 6.8% pectins and 6.3% lignols. Thus, the obtained material is a biocomposite, comprised of a heterogeneous mixture of naturally synthesized biopolymers. TGA curves of the biocomposite reveal four distinct mass loss steps (Fig. 2F). The first derivative of mass loss (DTG) peaks correspond to: evaporation of bound water (peak 1), and degradation of pectins (peak 2), hemicelluloses (peak 3), cellulose (peak 4), and phenolic compounds (peak 5-6)(*24*). The char residue is 10±5 wt%. The XRD patterns reveal multiple polymorphs of semicrystalline cellulose (I, II and III, marked in Fig. 2G) (*25*). Native cellulose from plant species crystallizes in the type I polymorphs. In our dehydrated biocomposites, cellulose microfibrils partially undergo phase transformations into crystal structures II and III, likely in response to the changing chemical environment during cell dissociation, the pressure applied during dehydration and the post-fabrication thermal treatment.

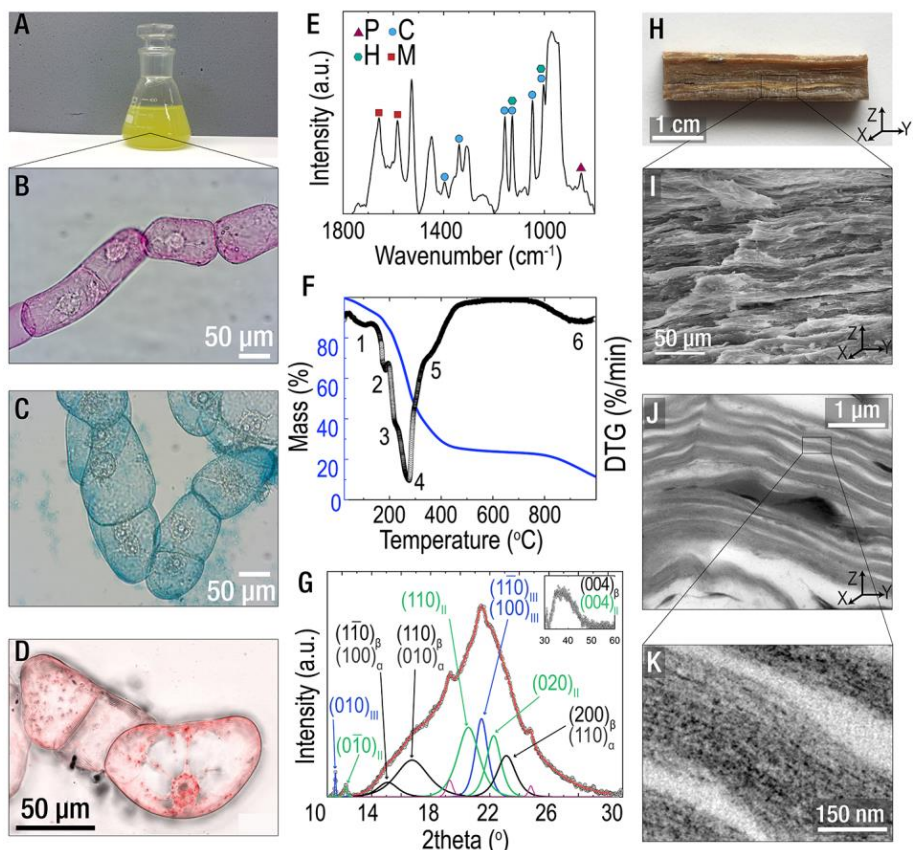

**Fig. 2.** (**A**) Photograph of the cell culture. Microscopy images of the cells stained for (**B**) pectins, (**C**) cellulose, and (**D**) lignols. (**E**) Raman spectrum of plant cells; peaks assigned to pectin (*P*), cellulose (*C*), hemicellulose (*H*) and monolignols (*M*). (**F**) TGA (blue line) and DTG (black dots) plots of the dehydrated biocomposite. (**G**) XRD pattern with marked contributions from cellulose polymorphs Iα, Iβ, II and III. (**H**) Photograph of the biocomposite sample. (**I**) SEM image of a cross-section, demonstrating the lamellar micro-structure. (**J**) TEM and (**K**) HRTEM images of the biocomposite cross-sections.

Optical and SEM observations of the biocomposites reveal an anisotropic, dense, lamellar microstructure comprised of compacted plant cells (Fig. 2H-I). Transmission electron microscopy (TEM) demonstrates that the primary cell walls are preserved during cell compression and dehydration (Fig. 2J-K). Accepted models suggest that the primary cell wall is a multi-lamellated structure consisting of cellulose microfibrils, arranged in various orientations within each plane (from entirely isotropic to somewhat aligned, depending on cell type), bound in a matrix of hemicelluloses and pectins (26). Even in the case of randomly distributed cellulose microfibrils in the plane of the wall, the structure is considered highly anisotropic across thickness (26). TEM images of our biocomposites show an average dehydrated cell wall thickness of 185±57 nm, and cellulose microfibrils diameters ranging between 1 and 30 nm. High resolution TEM (HRTEM) images confirm the presence of multi-lamellated structures, with cellulose microfibrils laying across the consecutive parallel planes (Fig. 2K, 3A-B). Using 3D tomographic reconstructions, we analyze the spatial distribution of the cell wall components, and observe their fibrous organization across multiple parallel planes, resulting in a highly anisotropic network (Fig. 3C). We observe a hierarchical microstructure: at the cellular level, a lamellar architecture consisting of compacted cells (Fig. 2I); at the sub-cellular level, an anisotropic, multi-lamellated structure, derived from the natural organization of the cell wall components (Figs. 2J-K, 3B-C).

We perform tensile and 3-point bending tests to characterize the mechanical performance of the dehydrated biocomposites. We compare them to different softwoods (pine), hardwoods (poplar, oak, and walnut), commercial plywood and MDF, and synthetic plastics of similar density (polystyrene, PS, polypropylene, PP, and low-density polyethylene, LDPE) (Fig. 3D-E, Fig. S1). Stress-strain plots obtained from the biocomposites (Fig. S2), show an initial linear elastic response upon loading, both under tension and bending, followed by a brittle failure at small strains (1± 0.3%). The Young's modulus, calculated from the initial linear elastic part of the tension experiments, is 2.5 ± 0.4 GPa, and the ultimate strength is 21.2 ± 3 MPa. The flexural modulus is 4.2 ± 0.4 GPa, and the modulus of rupture is 49.3 ± 3.2 MPa. Testing the flexural properties of the biocomposite on the two perpendicular planes (see schematic in Fig. S3), reveals that stiffness varies by a factor of ca. 1.75 in the two directions, while strength remains unaffected by orientation. The measured difference in stiffness is due to the anisotropic micro-structure of the biocomposite, resulting from the fabrication process which orients the cells normal to the compression direction.

We compare the mechanical properties of different woods and plastics (Fig. 3D-E, Fig. S1). Tension tests show that our biocomposites are stiffer than the other materials (Fig. 3D). However, natural woods have higher strength (Fig. 3E), which can be explained by their different cellular architectures, cell wall compositions, and components arrangements within the secondary cell walls. The cells used in our biocomposites originate from the herbaceous plant *Nicotiana tabacum* and they naturally develop a thin, unlignified primary cell wall (we detect only a low monolignol amount of 6.2 wt%). These cells do not form secondary cell walls and cannot self-organize in a hierarchical micro-structure in our cultures. Regardless, the mechanical performance of our biocomposites is comparable to that of commercial engineered woods and plastics. They surpass all literature-reported values for materials composed of plant cells, mycelium, or yeast matrixes (11, 14, 27) (Fig. 3G).

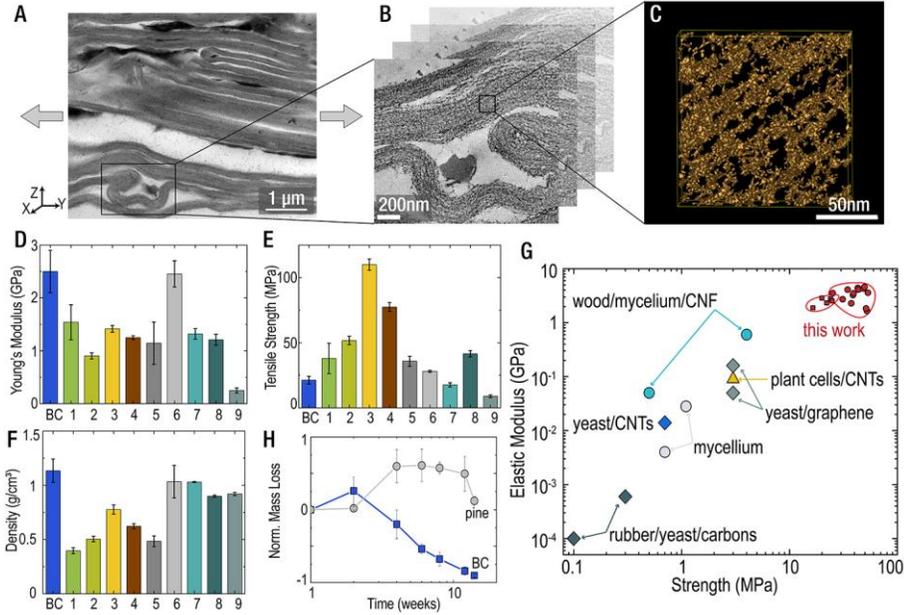

**Fig. 3. (A)** TEM of a cross-sectional area of the biocomposite. Gray arrows indicate the testing direction for tensile experiments. **(B)** Selected subsection for tomography imaging. **(C)** 3D reconstruction of selected cell wall subsection. Gold corresponds to dark pixels in the TEM, showing how the cell wall material is distributed in the selected area. **(D)** Young's modulus and **(E)** tensile strength of the biocomposite and reference materials. **(F)** Materials density. **(G)** Comparison of mechanical properties of this work (red points), and literature-reported biocomposites (*10–12*, *14*, *17*, *27*). **(H)** Biodegradation of the biocomposite and natural pine. Samples notation: BC: pure (without fillers) biocomposite; 1: pine; 2: poplar; 3: oak; 4: walnut; 5: plywood; 6: MDF; 7: PS; 8: PP; 9: LDPE.

A key factor in the design of sustainable materials is their end-of-life fate. The realization of biological matrix materials, such as those described here, offers an environmentally friendly alternative to non-degradable materials, which typically survive in landfills. To assess the biodegradability of our plant-based biocomposites, we perform agricultural soil incubation tests (see Methods), comparing their mass loss with that of natural wood (*28*). Results show an initial mass gain corresponding to water uptake from the soil, in both natural wood and biocomposites (Fig. 3H). The detectable mass loss due to biodegradation of the biocomposites begins 3 weeks after incubation, while for natural wood it begins about 7 weeks later. This can be associated to the presence of lignin in natural wood, which is known to provide resistance to pathogen attacks on cell walls (*29*). We observe an almost complete biodegradation of the biocomposite 14 weeks after initial incubation.

The proposed fabrication method allows us to use the natural biopolymer mixture as a matrix and incorporate filler additives, which (i) introduces new properties/functions in the composites, and (ii) enables further tuning of the mechanical performance. The addition of different amounts of carbon fibers (CF), for example, changes the biocomposites' compressive modulus and strength (Fig. 4A). For CF concentrations below 5 wt% there is a gradual improvement of elastic modulus and strength, followed by a decrease for higher concentrations, as observed in polymer composites because of fillers' aggregation (*30*). Different filler particles expand the biocomposites' property space (Fig. 4B). We plot the elastic modulus as a function of density of different plant-based biocomposites: pure cell matrix (BC), biocomposites containing various amounts of CF, halloysite and montmorillonite nanoclays (NC) and graphene (G). Their properties lie at the intersection of natural cellular materials and commercial plastics (Fig. 4B), presenting elastic moduli spanning over one order of magnitude. Filler additives also endow new functionalities, such as electrical conductivity or magnetic properties. The electrical conductivity of plant cell/CF composites, for example, can be tuned varying the CF content (Fig. 4C). Similarly, the addition of 13.5 wt% iron oxide nanoparticles (IN) in the plant cell matrix conveys ferro-magnetic properties, which allow the biocomposite to support more than five times its weight when attracted by a magnet (Fig. 4D).

We have developed a new method to create natural biocomposite materials based on plant cells. The method capitalizes on the plant cell's ability to synthesize intricate multi-lamellated structures of cellulose, hemicellulose, lignin and pectin in their cell walls. In the future, the use of different cell cultures and/or genetically modified species may allow the fabrication of materials with significantly altered properties. Similar fabrication approaches can be envisioned for many other biological systems (e.g. algae, fungi, etc.) that can provide complex elements as building blocks for advanced composite biomaterials.

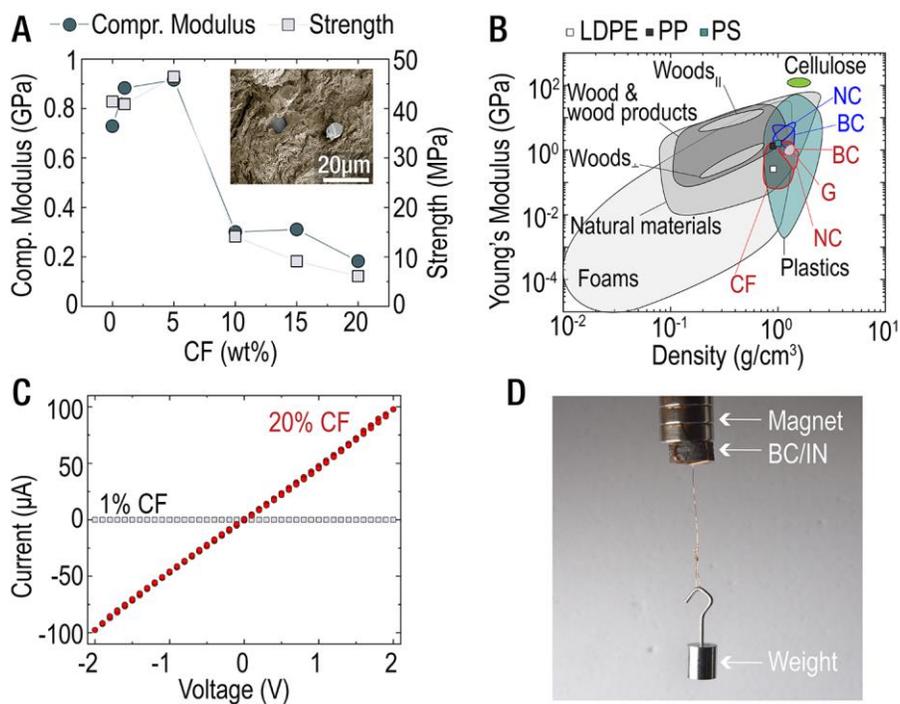

**Fig. 4.** (**A**) Compressive modulus and strength of biocomposites with CF. Inset: SEM image of the biocomposite with 1wt% CF (false colored). (**B**) Young's modulus versus density for various materials and our biocomposites. Blue groups correspond to bending experiments, red groups to compression. (**C**) IV curves for biocomposites with 1wt% and 20wt% CF. (**D**) Biocomposite with IN exhibiting magnetic properties.

**Acknowledgments:** Mr. Mark Ladinsky, Dr. Stavros Amanatidis, Dr. Yuchen Wei, Dr. Michael Mello, Mr. Azhar Carim, Ms. Sarah Antilla and Mr. Duy Anh Nguyen for support in experiments. Prof. Nathan Lewis for providing access to the Raman facilities. Caltech Kavli Nanoscience Institute, Gordon and Betty Moore, and the Beckman Foundation for gifts to Caltech to support electron microscopy. The Caltech Beckman Institute and the Arnold and Mabel Beckman Foundation for supporting the laser scanning imaging facilities.

**Funding:** L.B. was supported by the Swiss National Science Foundation under the Early Postdoc.Mobility project P2EZP2_175157. K.R. and the Raman microscope were supported by the Joint Center for Artificial Photosynthesis, a DOE Energy Innovation Hub, supported through the U.S. Department of Energy Office of Science under award number DE-SC0004993.

**Author contributions:** L.B., E.R., and C.D. conceived and directed the study. L.B. and E.R. designed the experiments. E.R. performed and oversaw all experiments, curated and analyzed the data, generated the figures. L.B. fabricated and tested biocomposites with carbon fibers. R.H. oversaw plant cell cultures, performed biodegradation experiments and optical microscopy. K.R. performed Raman measurements. All authors discussed the data. E.R. and C.D. wrote the manuscript with help from all authors.

**Competing interests:** Authors declare no competing interests.

**Data and materials availability:** All data in the main text and supplementary material will be made available upon request to the corresponding author.